\documentclass{article}

\usepackage[preprint]{neurips_2026}
\makeatletter
\renewcommand{\@noticestring}{}
\makeatother

\usepackage[utf8]{inputenc} 
\usepackage[T1]{fontenc}    
\usepackage{hyperref}       
\usepackage{url}            
\usepackage{booktabs}       
\usepackage{amsmath}        
\usepackage{amsfonts}       
\usepackage{nicefrac}       
\usepackage{microtype}      
\usepackage{xcolor}         
\usepackage{float}          
\usepackage{graphicx}       
\usepackage{pgfplots}       
\pgfplotsset{compat=1.18}
\usepgfplotslibrary{fillbetween}   
\usetikzlibrary{arrows.meta, positioning, shapes.geometric, backgrounds, calc, fit}

\definecolor{oraclegold}{HTML}{FFC107}
\definecolor{consensusblue}{HTML}{009999}
\definecolor{randomred}{HTML}{F44336}
\definecolor{topopurple}{HTML}{7B1FA2}

\title{Test-Time Scaling for CAD Generation via Verifier-Free Consensus Selection}

\author{%
  Aaron Haag \\
  \texttt{aaron.haag@siemens.com}
  \And
  Altay Kaçan \\
  \texttt{altay.kacan@siemens.com}
  \And
  Bertram Fuchs \\
  \texttt{bertram.fuchs@siemens.com}
  \And
  Oliver Lohse \\
  \texttt{oliver.lohse@siemens.com}
  \AND
  {\normalfont Siemens AG, Foundational Technologies, Munich, Germany} \\
}

\begin{document}

\raggedbottom

\maketitle

\begin{abstract}
Large language models can write parametric CAD programs from a natural-language
description (text-to-CAD generation), but a single sample is often wrong. Increasing test-time compute
by sampling multiple candidates only helps if a good candidate can be identified, yet
no ground-truth model is available at generation time. Existing systems often require a separate
verifier, such as a vision-language judge, to select among candidates. We
investigate whether the candidate pool itself provides enough signal for
effective selection and a verifier-free alternative. We introduce \emph{3D CAD consensus selection}, hereafter
\emph{consensus selection}:
sample $N$ parametric CAD programs, compile them to 3D models, and return the
candidate that agrees most with the rest of the pool. The method is
training-free and compatible with existing CAD agents.
We investigate geometric and topological notions of agreement, each of which
improves its corresponding evaluation metric. On the exact candidate pools of a
state-of-the-art CAD generation method, geometric consensus improves all three
geometric metrics over the method's verifier, while topological consensus
matches it on topology. Across every tested LLM and prompt variant, geometric
consensus also improves geometric accuracy over random selection from the same
pool, reducing Chamfer distance by 1--10\%.
\end{abstract}

\section{Introduction}
\label{sec:introduction}

Large language models can write parametric CAD programs from a natural-language
description \citep{khan2024text2cad,cadcodeverify,guan2025cadcoder}, for example
as executable parametric CAD code~\citep{cadquery}. A single sample, however, is often
wrong: dimensions are inaccurate, features are missing, or the compiled shape does not
match the prompt. A common mitigation strategy is to sample multiple candidates and select the most viable one.
The open question is how to pick when there is no ground-truth model, no
learned reward model, and no feedback loop that judges each candidate.

Recent CAD systems use vision-language or reasoning models to evaluate and
select generated candidates \citep{cadcodeverify,evocad}. This introduces an
additional model-dependent judgment whose decisions can be ambiguous or
inconsistent. We therefore test whether agreement among generated 3D CAD models
is sufficient for selection.

Our method is based on \emph{consensus selection}: we sample $N$ programs from an
LLM, compile them to 3D models, and select the candidate that agrees most with the
rest of the pool. The idea that agreement between samples is a useful signal is
well established for text and code \citep{kumar2004mbr,wang2022selfconsistency,shi2022mbrexec}.
CAD provides a natural domain for consensus because each program compiles into
an explicit 3D model. We define agreement using properties of these models,
focusing on geometry and topology.

Our contributions are:
\begin{itemize}
  \item We introduce consensus selection, a verifier-free inference-time
  selection method for LLM-based parametric 3D CAD generation. It defines
  consensus over properties of compiled 3D CAD models, specifically geometry
  and topology. It is training-free, requires no feedback loop, and can be
  integrated into existing CAD generation pipelines.
  \item Using the exact candidate pools of a state-of-the-art verifier-based
  CAD generation method, consensus selection improves all evaluated geometric
  metrics over the method's verifier, while matching it on topology.
  \item We study test-time scaling with candidate-pool size and show how
  selection gains vary across LLMs and prompt variants.
\end{itemize}

\section{Related work}
\label{sec:related-work}

\paragraph{Text-to-parametric-CAD generation.}
Early systems encode construction histories as specialized command sequences
\citep{wu2021deepcad}; Text2CAD made them text-conditioned
\citep{khan2024text2cad}, followed by more controllable or language-driven
variants \citep{zhang2025flexcad,li2025cadllama,wang2025cadfusion}. Code-based
methods instead exploit pretrained LLMs by emitting executable parametric CAD
code~\citep{cadquery}. CadCodeVerify introduced the CADPrompt benchmark
\citep{cadcodeverify}; Text-to-CadQuery \citep{xie2025texttocadquery} and
CAD-Coder \citep{guan2025cadcoder} fine-tune for this representation, and
FutureCAD grounds textual references to B-Rep primitives
\citep{li2026futurecad}. Newer benchmarks test geometric complexity
\citep{wang2026text2cadbench}, engineering criteria \citep{dong2026muse}, and
drawing-conditioned generation and editing \citep{cadgenbench2026}. We use
CADPrompt throughout. Reconstruction methods can rank candidates against an
input point cloud, mesh, or drawing
\citep{rukhovich2025cadrecode,kolodiazhnyi2025cadrille}, and vision-language
models can emit CadQuery programs directly from images \citep{doris2026cad};
our text-only setting has no such reference.

\paragraph{Verifier-based refinement and ranking.}
Feedback enters CAD generation either during training
\citep{wang2025cadfusion,giannone2026gift} or at inference time. Most
inference pipelines iteratively refine one program using rendered views
\citep{yuan2024-3dpremise,li2026seekcad}, reviewer or tool-using agents
\citep{panta2025meda,shui2026articad,barkley2026cadsmith,mallis2025cadassistant}, or compiler,
geometric, and simulation feedback
\citep{badagabettu2024query2cad,zhou2026cadialogue,son2026fea,berger2026physics,liu2026embodiedcad}.
CadCodeVerify, a baseline in Section~\ref{sec:exp-verifier}, has a VLM generate
and answer validation questions about a render. Closer to our setting, EvoCAD
\citep{evocad} evolves a
population using VLM descriptions and a reasoning-model ranker, motivated by
evidence that VLMs can assess 3D assets well \citep{wu2024gptv-eval}. Such judges
add model calls and can exhibit self-preference
\citep{zheng2023judge,panickssery2024selfpreference}, weak discrimination
\citep{west2023paradox}, over-optimization \citep{gao2023overoptimization}, or
gaming \citep{zhou2026cadjudge}. CAD-Judge \citep{zhou2026cadjudge} instead uses
the compiler to score Chamfer distance to a ground-truth solid, but therefore
requires a reference unavailable at generation time. We use only agreement
among the candidates.

\paragraph{Minimum Bayes risk decoding over parallel samples.}
Our selection rule is an instance of minimum Bayes risk (MBR) decoding
\citep{kumar2004mbr}: instead of the most probable output, MBR returns the
candidate with the lowest expected loss against the output distribution, here
approximated by the empirical distribution over the drawn samples
\citep{eikema2022sampling,bertsch2023mbr}. The loss determines which kind of
agreement is rewarded: a 0-1 loss over an equivalence relation reduces MBR to
majority voting, as in self-consistency \citep{wang2022selfconsistency}, while a
continuous metric returns the medoid of the pool. In program synthesis, program
text is a poor equivalence class, since many distinct programs implement the same
function. MBR-EXEC \citep{shi2022mbrexec} resolves this by executing each
sample and voting on execution results, and AlphaCode \citep{li2022alphacode}
clusters candidates by execution behavior; universal self-consistency
\citep{chen2023usc} instead delegates the grouping to an LLM, reintroducing a
language-model judge. Search-based test-time compute \citep{yao2023tot} is
orthogonal: consensus selection applies to whatever pool it produces.

We take the same step as MBR-EXEC, with the CAD compiler as the interpreter:
executing a program yields a 3D model, and agreement is measured over properties
of that model rather than over program text or a VLM's description of it.
Compiled CAD models support both continuous geometric and discrete topological
comparisons. We therefore instantiate both regimes, namely a continuous
Chamfer loss giving a geometric medoid, and a 0-1 loss over
Euler-characteristic classes giving a coarse topology-derived majority vote.
Empirically, each performs best on the
evaluation metric most closely aligned with its loss, consistent with the
objective dependence of MBR. This mirrors MBR-EXEC's use of execution
semantics for code. The gap
between our selection and an oracle over the same pool is analogous to the
coverage-versus-selection gap in repeated sampling \citep{brown2024monkeys}:
candidate pools often contain better solutions than an agreement-based rule
recovers.

\section{Method}
\label{sec:method}

In this section, we formally define the task of generating parametric 3D-CAD
models from natural-language prompts and introduce \emph{3D CAD consensus selection},
a verifier-free method for selecting a final model from a set of LLM-generated
candidates. Figure~\ref{fig:teaser} gives an overview: given a prompt, we
sample $N$ candidate CAD programs from an LLM, compile them into 3D models,
and select the candidate that agrees most with the rest of the pool.

\newcommand{\cadpart}[5]{%
  \begin{scope}[shift={(#1,#2)}, scale=#3, x={(-0.5cm,-0.25cm)}, y={(0.5cm,-0.25cm)}, z={(0cm,0.55cm)}, line join=round, line cap=round, line width=0.4pt]
    \def\w{1.0} \def\d{1.0} \def\h{0.4}
    \ifnum#5=2 \def\h{0.15}\fi 
    \ifnum#5=3 \def\w{0.6}\fi  
    \filldraw[fill=#4!60, draw=#4!75!black] (0,0,0) -- (\w,0,0) -- (\w,0,\h) -- (0,0,\h) -- cycle;
    \filldraw[fill=#4!40, draw=#4!75!black] (\w,0,0) -- (\w,\d,0) -- (\w,\d,\h) -- (\w,0,\h) -- cycle;
    \filldraw[fill=#4!15, draw=#4!75!black] (0,0,\h) -- (\w,0,\h) -- (\w,\d,\h) -- (0,\d,\h) -- cycle;
    \ifnum#5=5 
    \else
      \def\hx{0.5*\w} \def\hy{0.5*\d}
      \ifnum#5=4 \def\hx{0.2*\w} \def\hy{0.2*\d}\fi
      \fill[white, draw=#4!75!black, line width=0.4pt] (\hx,\hy,\h) ellipse [x radius=0.18, y radius=0.09];
    \fi
  \end{scope}
}

\begin{figure}[H]
  \centering
  \resizebox{\linewidth}{!}{%
  \begin{tikzpicture}[
    font=\footnotesize,
    sub/.style={font=\bfseries\footnotesize, color=black!90, align=center},
    box/.style={draw=black!30, rounded corners=3pt, fill=black!2, inner sep=6pt, line width=0.6pt},
    flow/.style={-{Stealth[length=5pt]}, line width=1pt, color=black!35},
  ]
    \def\ty{1.3} 
    
    \node[box, text width=2.2cm] (p1) at (0,0) {``L-shaped bracket with a centered slotted hole\ldots''};
    \node[sub]   at (0,\ty)  {Natural language};

    \begin{scope}[shift={(3.5,0)}]
      \foreach \o in {0.16, 0.08, 0}
        \filldraw[fill=white, draw=black!35, line width=0.6pt, rounded corners=1.5pt]
          (-0.55+\o,-0.65+\o) rectangle ++(1.1,1.3);
      \foreach \y in {0.4, 0.2, 0, -0.2, -0.4}
        \draw[consensusblue!70, line width=1pt, line cap=round] (-0.4, \y) -- ++(0.7,0);
    \end{scope}
    \node[sub]   at (3.5,\ty)  {$N$ parametric CAD programs};

    \begin{scope}[shift={(7.0,0)}]
      \node[draw=black!15, dashed, rounded corners=4pt, minimum width=2.4cm, minimum height=2.1cm] (b3) at (0,0) {};
      \cadpart{-0.45}{0.475}{0.4}{gray}{2}
      \cadpart{0.45}{0.475}{0.4}{gray}{3}
      \cadpart{0.0}{0.05}{0.5}{consensusblue}{1}
      \cadpart{0.45}{-0.375}{0.4}{gray}{5}
      \cadpart{-0.45}{-0.375}{0.4}{gray}{4}
    \end{scope}
    \node[sub]   at (7.0,\ty)  {3D CAD model pool};

    \begin{scope}[shift={(10.5,0)}]
      \def\sz{0.24}
      \def\mx{-0.795} 
      \foreach \i/\j/\v in {0/1/55, 0/2/26, 0/3/72, 0/4/46, 1/2/20, 1/3/62, 1/4/50, 2/3/30, 2/4/16, 3/4/66}{
        \fill[consensusblue!\v] ({\mx+\j*\sz}, {0.6-\i*\sz}) rectangle ++(\sz,-\sz);
        \fill[consensusblue!\v] ({\mx+\i*\sz}, {0.6-\j*\sz}) rectangle ++(\sz,-\sz);
      }
      \foreach \i in {0,...,4}{
        \fill[white] ({\mx+\i*\sz}, {0.6-\i*\sz}) rectangle ++(\sz,-\sz);
      }
      \draw[black!40, line width=0.5pt] (\mx,-0.6) rectangle ++(1.2,1.2);
      \def\ax{0.555} 
      \foreach \i/\v in {0/49, 1/37, 2/18, 3/57, 4/44}{
        \fill[consensusblue!\v] (\ax, {0.6-\i*\sz}) rectangle ++(\sz,-\sz);
      }
      \draw[black!40, line width=0.5pt] (\ax,-0.6) rectangle ++(\sz,1.2);
      \node[font=\tiny, color=black!60] at ({\mx+0.6}, 0.75) {$d_{ij}$};
      \node[font=\tiny, color=black!60] at ({\ax+0.12}, 0.75) {$\bar{d}_i$};
      \draw[consensusblue, line width=1.1pt, rounded corners=1.5pt]
        ({\mx-0.04}, {0.6-2*\sz+0.02}) rectangle ({\ax+\sz+0.04}, {0.6-3*\sz-0.02});
      \node[font=\scriptsize\bfseries, color=consensusblue] at (0, -0.9) {$\arg\min_i \bar{d}_i$};
    \end{scope}
    \node[sub]   at (10.5,\ty)  {Consensus selection};

    \begin{scope}[shift={(14.0,0)}]
      \node[draw=consensusblue, fill=consensusblue!5, rounded corners=3pt, minimum width=1.8cm, minimum height=1.8cm, line width=1pt] (o5) at (0,0) {};
      \cadpart{0.0}{0.11}{0.8}{consensusblue}{1}
    \end{scope}
    \node[sub, color=consensusblue!80!black] at (14.0,\ty) {Final CAD model};

    \draw[flow] (1.57,0) -- ++(0.9,0); 
    \draw[flow] (4.47,0) -- ++(0.9,0); 
    \draw[flow] (8.52,0) -- ++(0.9,0); 
    \draw[flow] (11.75,0) -- ++(0.9,0); 
  \end{tikzpicture}}%
  \caption{\textbf{Consensus selection for 3D CAD generation.} We compile $N$
  sampled parametric CAD programs into 3D models and select one based on
  agreement in their 3D CAD model properties.}
  \label{fig:teaser}
\end{figure}
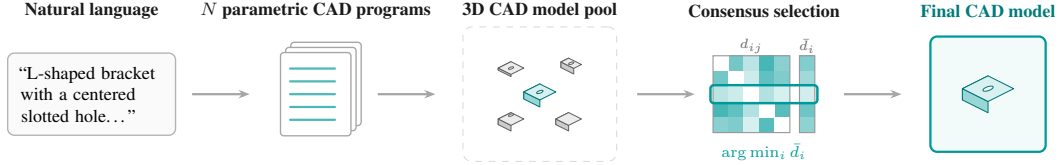

\paragraph{Problem setting.}
The task is to generate a parametric CAD program $c$ from a natural-language
prompt $p$; a compiler $\phi$ converts the program into a 3D model $\phi(c)$
\citep{evocad} facilitating a direct comparison of properties of 3D models.
The ground-truth object is not available at generation time, and the only input
is the prompt. Instead of generating a single program, we sample $N$ candidate
programs $c_1, \dots, c_N$ independently from an LLM conditioned on $p$
(Section~\ref{sec:experiments} details our setup). Candidates that fail to
compile are discarded, leaving a set of valid 3D models
$\{S_i = \phi(c_i) \mid i \in V\}$ with $V \subseteq \{1, \dots, N\}$. The
goal is to select one model from this pool.

\paragraph{Consensus selection.}
This method is based on the principle that correct features appear in many candidates, while individual mistakes are random. We therefore select the candidate that best agrees with the group. Specifically, for a distance function $d$, we choose the model with the lowest average distance to all others,
\begin{equation}
  \label{eq:medoid}
  i^{*} \;=\; \arg\min_{i \in V} \; \frac{1}{|V| - 1}
  \sum_{j \in V \setminus \{i\}} d\!\left(S_i, S_j\right).
\end{equation}
We define this decision rule as \emph{consensus selection} yielding the respective \emph{consensus candidate}. Mathematically, this is minimum Bayes risk decoding with $d$ as the loss function and a uniform distribution over the sampled candidates \citep{kumar2004mbr,eikema2022sampling}. We apply this principle to continuous 3D geometry, where exact-match voting \citep{shi2022mbrexec} is not possible. The choice of $d$ determines what kind of agreement is rewarded. We use both geometric and topological distances.

\paragraph{Geometric consensus.}
For each model $S_i$ we sample a point cloud $P_i \subset \mathbb{R}^3$
uniformly from its surface; each pair of models is normalized and aligned with
ICP beforehand. Geometric agreement between two models is measured by the
symmetric Chamfer distance between their point clouds,
\begin{equation}
  \label{eq:chamfer}
  d_{\mathrm{CD}}(S_i, S_j) \;=\;
  \frac{1}{2}\left(
  \frac{1}{|P_i|} \sum_{x \in P_i} \min_{y \in P_j} \lVert x - y \rVert_2
  \;+\;
  \frac{1}{|P_j|} \sum_{y \in P_j} \min_{x \in P_i} \lVert x - y \rVert_2
  \right),
\end{equation}
where $x$ and $y$ are points of $P_i$ and $P_j$. With
$d = d_{\mathrm{CD}}$, Equation~\eqref{eq:medoid} picks the candidate with the
lowest average Chamfer distance to all other candidates, i.e., the
geometrically most central model of the pool.

\paragraph{Topological consensus.}
We also compare candidates by their topology, using the Euler characteristic
$\chi(S)$ of the mesh (vertices minus edges plus faces; defined for watertight
meshes), following EvoCAD \citep{evocad}, who used it to evaluate the topology of
generated CAD models.
The topological distance between two models is
\begin{equation}
  \label{eq:topo}
  d_{\chi}(S_i, S_j) \;=\;
  \begin{cases}
    0 & \text{if } \chi(S_i) = \chi(S_j), \\
    1 & \text{otherwise.}
  \end{cases}
\end{equation}
With $d = d_{\chi}$, Equation~\eqref{eq:medoid} (computed over the
watertight candidates) selects the candidate from the largest group of
models with identical topology. Ties within this group are broken by the
geometric consensus score, i.e., by Equation~\eqref{eq:medoid} with
$d_{\mathrm{CD}}$.

\paragraph{Properties.}
Consensus selection is training-free and model-agnostic, it requires only a
compiler and $\mathcal{O}(|V|^2)$ pairwise distance evaluations, and it makes
no assumption about how the candidate pool was generated. It can therefore be
applied on top of any existing generation pipeline; in
Section~\ref{sec:experiments} we apply it directly to the candidate
populations of a verifier-based method.

\section{Experiments}
\label{sec:experiments}

We evaluate consensus selection in three steps: a comparison with
verifier-based selection on the original data of \citet{evocad}
(Section~\ref{sec:exp-verifier}), the effect of the number of sampled
candidates $N$ (Section~\ref{sec:exp-scaling}), and the behavior across
different LLMs and prompt variants (Section~\ref{sec:exp-models}).

\subsection{Experimental setup}
\label{sec:exp-setup}

\paragraph{Benchmark.}
All experiments are based on the CADPrompt benchmark \citep{cadcodeverify},
which contains 200 CAD objects with ground-truth models. Each object has two
natural-language descriptions (one with explicit measurements and one
without).

\paragraph{Candidate generation.}
For the scaling and model-comparison experiments we generate candidate pools
with the few-shot setup of EvoCAD \citep{evocad}: the LLM receives a short
CadQuery introduction, five programs drawn at random from EvoCAD's set of
40 few-shot examples, and the object description. Sampling uses temperature
0.5, and the few-shot examples are redrawn for every candidate.
Generated programs are compiled to meshes, and candidates that fail to compile
are discarded without any repair or self-debugging attempt. We generate 30
candidates per prompt for the models listed in Table~\ref{tab:model-comparison}.

\paragraph{Verifier-based comparison data.}
For the comparison with verifier-based methods, we apply consensus selection
directly to the original run data of \citet{evocad}. The initial populations
in this dataset contain six candidates per prompt and are shared across all
selection methods. This ensures that any difference in results stems only
from the selection strategy, as the underlying candidate pool remains
identical.

\paragraph{Metrics.}
We evaluate using Chamfer distance (CD), the
95th-percentile Hausdorff distance (HDD), and voxel IoU. For voxel IoU, we
sample 10{,}000 points from each aligned mesh and convert them into a solid
voxel grid with voxel size 0.1. We also report the absolute error in Euler
characteristic ($T_{err}$) and the percentage of correct topologies
($T_{corr}$). Finally, coverage is the percentage of prompts yielding at least
one successfully compiled candidate.

\subsection{Comparison with verifier-based selection}
\label{sec:exp-verifier}

Table~\ref{tab:main-results} compares consensus selection with the verifier of
\citet{evocad}. The comparison uses the exact same initial EvoCAD
populations. We also report the result after full evolution as a reference.

\paragraph{Protocol.}
For the comparison, we first restrict each initial EvoCAD population to its
watertight candidates. Random selection (computed as the mean over this set),
verifier selection, and both consensus variants are evaluated using these same
candidates. The verifier selects the highest-ranked candidate within the set.
This differs from the original EvoCAD protocol, in which the verifier could
select non-watertight candidates. Applying watertight filtering to all methods
that select from the initial population prevents topological consensus from
gaining an advantage because it excludes non-watertight meshes by definition.

The published full-evolution result is included only as a reference. Since
evolution had already been completed in the available EvoCAD data, we could
not apply watertight filtering at selection time. Full evolution also
generated and evaluated additional candidates over multiple evolutionary
generations, whereas the selection methods use only the initial candidate
pools.

The CADPrompt benchmark contains 200 prompts. Compiled meshes are available
for 197 prompts in the CadCodeVerify supplementary material. To compare all
methods on the same prompts, geometry is therefore evaluated on these 197
prompts. Topology metrics are only defined for watertight meshes. Topology is
therefore evaluated on the subset of 175 prompts for which these metrics are
available for every method. In Table~\ref{tab:main-results}, $n$ denotes the
number of prompts used.

\paragraph{Geometry.}
Both consensus variants improve all geometry metrics over verifier selection
from the same candidate pool. Geometric consensus improves CD, HDD, and IoU
over the verifier (all $p\leq0.022$). Topological consensus significantly
improves CD and IoU over verifier-based selection ($p=0.003$ and $p=0.035$).
Both consensus variants also obtain lower CD and higher IoU than full
evolution.

\paragraph{Topology.}
Topological consensus improves over random selection ($T_{err}$ 0.542 to
0.465; $T_{corr}$ 82.1\% to 84.9\%).
We detect no significant topology difference between topological consensus
and verifier-based selection ($p=0.82$ for $T_{err}$ and $p=1.00$ for
$T_{corr}$). Full evolution is shown as a reference and obtains numerically
better topology scores after generating additional candidates over multiple
evolutionary generations.

\begin{table}[t]
  \caption{Comparison with verifier-based methods on the CADPrompt benchmark
  \citep{cadcodeverify}. Bold denotes the best result among methods selecting
  from the same initial candidate pools. CadCodeVerify and full evolution are
  shown as references.
  }
  \label{tab:main-results}
  \centering
  \setlength{\tabcolsep}{4pt}
  \begin{tabular}{lccccc}
    \toprule
    & \multicolumn{3}{c}{Geometry ($n=197$)} & \multicolumn{2}{c}{Topology ($n=175$)} \\
    \cmidrule(lr){2-4}\cmidrule(lr){5-6}
    Method & CD $\downarrow$ & HDD $\downarrow$ & IoU $\uparrow$ & $T_{err}$ $\downarrow$ & $T_{corr}$ (\%) $\uparrow$ \\
    \midrule
    \multicolumn{6}{l}{\emph{Reference methods}} \\
    CadCodeVerify \citep{cadcodeverify}           & 0.0657 & 0.191 & 0.684 & 0.629 & 80.6 \\
    EvoCAD (full evolution) \citep{evocad}         & 0.0634 & 0.180 & 0.688 & 0.350 & 88.3 \\
    \midrule
    \multicolumn{6}{l}{\emph{Selection from shared watertight subsets of the initial pools}} \\
    Random selection            & 0.0631 & 0.183 & 0.693 & 0.542 & 82.1 \\
    EvoCAD verifier \citep{evocad} & 0.0627 & 0.181 & 0.695 & 0.484 & 84.8 \\
    Ours (geometric consensus)   & \textbf{0.0610} & \textbf{0.176} & \textbf{0.703} & 0.518 & 83.5 \\
    Ours (topological consensus) & 0.0611 & 0.177 & 0.700 & \textbf{0.465} & \textbf{84.9} \\
    \bottomrule
  \end{tabular}
\end{table}

\subsection{Scaling with the number of sampled candidates}
\label{sec:exp-scaling}

We next study how the benefit of consensus selection depends on the number of
sampled candidates $N$. For this experiment we use the gpt-oss-20b pools on
the prompts without explicit measurements (182 prompts for geometry metrics
and 180 with a watertight ground truth for $T_{corr}$). For each $N$, we sample
1000 candidate subsets of size $N$, resampling subsets without any watertight
candidate. Every method is restricted to the watertight candidates within each
sampled subset, as topological consensus would otherwise have an advantage by
inherently selecting only watertight meshes. Random selects uniformly, whereas
the oracle selects the best candidate for the considered metric.

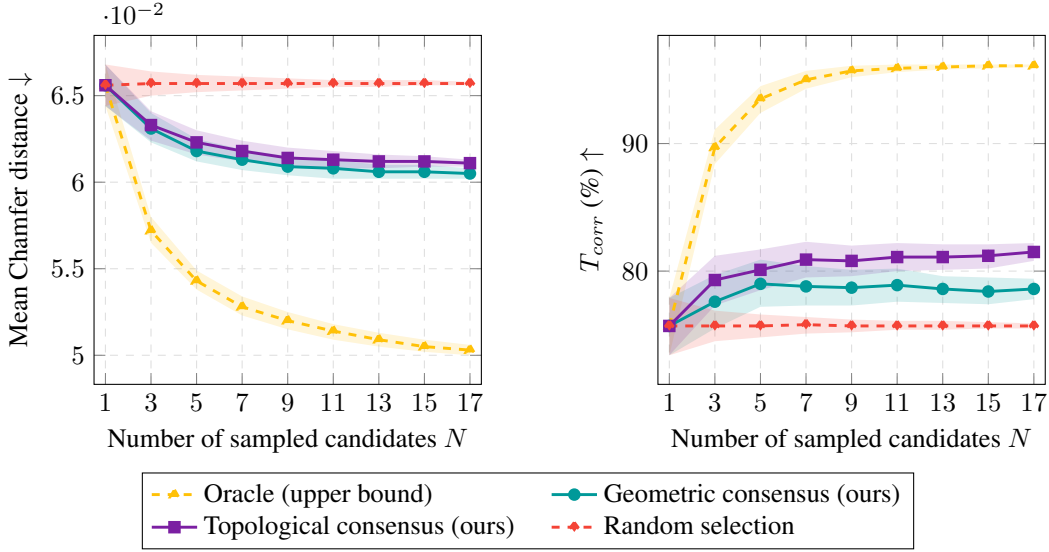
\begin{figure}[H]
  \centering
  \begin{tikzpicture}
    \begin{axis}[
        width=0.48\linewidth,
        height=6.2cm,
        xlabel={Number of sampled candidates $N$},
        ylabel={Mean Chamfer distance $\downarrow$},
        xmin=0.5, xmax=17.5,
        xtick={1,3,5,7,9,11,13,15,17},
        legend to name=scalinglegendwt,
        legend columns=2,
        legend style={/tikz/every even column/.append style={column sep=0.4cm}},
        legend cell align=left,
        grid=major,
        grid style={dashed, gray!25},
        every axis plot/.append style={line width=1.1pt, mark size=1.8pt},
      ]
      \addplot[forget plot, draw=none, fill=oraclegold, opacity=0.16] coordinates {
        (1,0.0668) (3,0.0580) (5,0.0549) (7,0.0534) (9,0.0525) (11,0.0518) (13,0.0513) (15,0.0509) (17,0.0506)
        (17,0.0500) (15,0.0502) (13,0.0505) (11,0.0509) (9,0.0515) (7,0.0523) (5,0.0538) (3,0.0565) (1,0.0644)
      } \closedcycle;
      \addplot[forget plot, draw=none, fill=consensusblue, opacity=0.16] coordinates {
        (1,0.0668) (3,0.0640) (5,0.0625) (7,0.0619) (9,0.0615) (11,0.0613) (13,0.0610) (15,0.0610) (17,0.0608)
        (17,0.0602) (15,0.0602) (13,0.0602) (11,0.0602) (9,0.0604) (7,0.0607) (5,0.0612) (3,0.0623) (1,0.0644)
      } \closedcycle;
      \addplot[forget plot, draw=none, fill=topopurple, opacity=0.16] coordinates {
        (1,0.0668) (3,0.0641) (5,0.0630) (7,0.0624) (9,0.0620) (11,0.0618) (13,0.0616) (15,0.0615) (17,0.0613)
        (17,0.0609) (15,0.0609) (13,0.0608) (11,0.0608) (9,0.0609) (7,0.0612) (5,0.0616) (3,0.0624) (1,0.0644)
      } \closedcycle;
      \addplot[forget plot, draw=none, fill=randomred, opacity=0.16] coordinates {
        (1,0.0668) (3,0.0664) (5,0.0662) (7,0.0661) (9,0.0660) (11,0.0659) (13,0.0659) (15,0.0659) (17,0.0658)
        (17,0.0656) (15,0.0656) (13,0.0655) (11,0.0655) (9,0.0654) (7,0.0653) (5,0.0652) (3,0.0650) (1,0.0644)
      } \closedcycle;
      \addplot[oraclegold, mark=triangle*, dashed] coordinates {
        (1, 0.0656) (3, 0.0572) (5, 0.0543) (7, 0.0528) (9, 0.0520)
        (11, 0.0514) (13, 0.0509) (15, 0.0505) (17, 0.0503)
      };
      \addlegendentry{Oracle (upper bound)}
      \addplot[consensusblue, mark=*] coordinates {
        (1, 0.0656) (3, 0.0631) (5, 0.0618) (7, 0.0613) (9, 0.0609)
        (11, 0.0608) (13, 0.0606) (15, 0.0606) (17, 0.0605)
      };
      \addlegendentry{Geometric consensus (ours)}
      \addplot[topopurple, mark=square*] coordinates {
        (1, 0.0656) (3, 0.0633) (5, 0.0623) (7, 0.0618) (9, 0.0614)
        (11, 0.0613) (13, 0.0612) (15, 0.0612) (17, 0.0611)
      };
      \addlegendentry{Topological consensus (ours)}
      \addplot[randomred, mark=diamond*, dashed] coordinates {
        (1, 0.0656) (3, 0.0657) (5, 0.0657) (7, 0.0657) (9, 0.0657)
        (11, 0.0657) (13, 0.0657) (15, 0.0657) (17, 0.0657)
      };
      \addlegendentry{Random selection}
    \end{axis}
  \end{tikzpicture}%
  \hfill
  \begin{tikzpicture}
    \begin{axis}[
        width=0.48\linewidth,
        height=6.2cm,
        xlabel={Number of sampled candidates $N$},
        ylabel={$T_{corr}$ (\%) $\uparrow$},
        xmin=0.5, xmax=17.5,
        xtick={1,3,5,7,9,11,13,15,17},
        grid=major,
        grid style={dashed, gray!25},
        every axis plot/.append style={line width=1.1pt, mark size=1.8pt},
      ]
      \addplot[forget plot, draw=none, fill=oraclegold, opacity=0.16] coordinates {
        (1,77.9) (3,91.1) (5,94.5) (7,95.7) (9,96.1) (11,96.2) (13,96.2) (15,96.2) (17,96.2)
        (17,96.0) (15,95.9) (13,95.8) (11,95.6) (9,95.2) (7,94.3) (5,92.4) (3,88.4) (1,73.4)
      } \closedcycle;
      \addplot[forget plot, draw=none, fill=consensusblue, opacity=0.16] coordinates {
        (1,77.9) (3,79.7) (5,80.9) (7,80.4) (9,80.1) (11,80.2) (13,79.6) (15,79.5) (17,79.4)
        (17,77.8) (15,77.4) (13,77.5) (11,77.6) (9,77.3) (7,77.3) (5,77.2) (3,75.5) (1,73.4)
      } \closedcycle;
      \addplot[forget plot, draw=none, fill=topopurple, opacity=0.16] coordinates {
        (1,77.9) (3,81.2) (5,81.7) (7,82.3) (9,82.0) (11,82.2) (13,82.1) (15,82.1) (17,82.2)
        (17,80.8) (15,80.2) (13,80.1) (11,80.0) (9,79.6) (7,79.5) (5,78.5) (3,77.3) (1,73.4)
      } \closedcycle;
      \addplot[forget plot, draw=none, fill=randomred, opacity=0.16] coordinates {
        (1,77.9) (3,76.9) (5,76.6) (7,76.4) (9,76.2) (11,76.1) (13,76.1) (15,76.0) (17,75.9)
        (17,75.5) (15,75.5) (13,75.4) (11,75.4) (9,75.2) (7,75.1) (5,74.8) (3,74.5) (1,73.4)
      } \closedcycle;
      \addplot[oraclegold, mark=triangle*, dashed] coordinates {
        (1, 75.7) (3, 89.7) (5, 93.5) (7, 95.0) (9, 95.7)
        (11, 95.9) (13, 96.0) (15, 96.1) (17, 96.1)
      };
      \addplot[consensusblue, mark=*] coordinates {
        (1, 75.7) (3, 77.6) (5, 79.0) (7, 78.8) (9, 78.7)
        (11, 78.9) (13, 78.6) (15, 78.4) (17, 78.6)
      };
      \addplot[topopurple, mark=square*] coordinates {
        (1, 75.7) (3, 79.3) (5, 80.1) (7, 80.9) (9, 80.8)
        (11, 81.1) (13, 81.1) (15, 81.2) (17, 81.5)
      };
      \addplot[randomred, mark=diamond*, dashed] coordinates {
        (1, 75.7) (3, 75.7) (5, 75.7) (7, 75.8) (9, 75.7)
        (11, 75.7) (13, 75.7) (15, 75.7) (17, 75.7)
      };
    \end{axis}
  \end{tikzpicture}

  \vspace{0.3em}
  \pgfplotslegendfromname{scalinglegendwt}

  \caption{Scaling with the number of sampled candidates $N$ (gpt-oss-20b,
  prompts without explicit measurements; 182 prompts for CD, 180 with
  watertight ground truth for $T_{corr}$). Left: mean Chamfer distance of the
  selected candidate (lower is better). Right: topology correctness $T_{corr}$
  of the selected candidate (higher is better). All methods select from the
  watertight candidates of each subset. Shaded bands show $\pm1$ standard deviation over the
  random subset draws.
  }
  \label{fig:scaling}
\end{figure}

Figure~\ref{fig:scaling} shows the result. Both consensus variants improve
over random selection already at $N{=}3$ and saturate quickly, with little
additional gain beyond $N \approx 9$. The oracle keeps improving
as the candidate pools contain better models than any agreement-based
selection recovers.

We observe that the choice of the selection criteria determines which metric
improves: geometric consensus performs best in Chamfer distance for every $N$,
while topological consensus achieves the highest $T_{corr}$. This shows that
consensus selection can be targeted to specific metrics by choosing the
corresponding dissimilarity.

\subsection{Consensus selection across models}
\label{sec:exp-models}

Finally, we apply the geometric consensus to the candidate pools of four
models, each with both CADPrompt prompt variants
(Table~\ref{tab:model-comparison}). We generate 30 candidates per prompt and
evaluate selection at $N{=}15$ with 1000 random subsets per prompt. To keep
the rows directly comparable, the CD columns are averaged over the common
subset of $n{=}125$ prompts on which every model has at least three compiling
candidates. Coverage and the number of compiling candidates are counted per
model over all prompts.

Consensus selection improves over random selection for every model and both
prompt variants, but the size of the gain varies: the relative CD improvement
is 8--10\% for Gemma 3 12B and gpt-oss-20b, and only 1--3\% for Gemini 3 Flash
and Gemma 4. The stronger models produce better random picks, and prompts with
explicit measurements generally lead to better geometry, albeit at slightly
lower coverage. In all settings the oracle remains clearly better than the
consensus candidate as it cannot exceed the best sampled
candidate by construction.

\begin{table}[H]
  \caption{Consensus selection across models and prompt variants at $N{=}15$.
  Cov.\ and Cand.\ denote compilation coverage and mean candidate count.
  }
  \label{tab:model-comparison}
  \centering
  \begin{tabular}{llccccc}
    \toprule
    Model & Prompt & Cov.\ (\%) & Cand. & Random $\downarrow$ & Ours $\downarrow$ & Oracle $\downarrow$ \\
    \midrule
    Gemma 3 12B & w/ measurements  & 87.5 & 13.3 & 0.0650 & 0.0596 & 0.0423 \\
    Gemma 3 12B & w/o measurements & 94.5 & 13.6 & 0.0658 & 0.0590 & 0.0446 \\
    \midrule
    gpt-oss-20b & w/ measurements  & 89.5 & 15.2 & 0.0562 & 0.0515 & 0.0409 \\
    gpt-oss-20b & w/o measurements & 97.0 & 16.1 & 0.0624 & 0.0572 & 0.0456 \\
    \midrule
    Gemma 4 & w/ measurements  & 98.5 & 22.3 & 0.0574 & 0.0557 & 0.0395 \\
    Gemma 4 & w/o measurements & 100.0 & 23.5 & 0.0616 & 0.0608 & 0.0458 \\
    \midrule
    Gemini 3 Flash & w/ measurements  & 97.0 & 26.1 & 0.0550 & 0.0546 & 0.0408 \\
    Gemini 3 Flash & w/o measurements & 99.0 & 22.7 & 0.0607 & 0.0592 & 0.0479 \\
    \bottomrule
  \end{tabular}
\end{table}

\section{Discussion and limitations}
\label{sec:discussion}

In Section~\ref{sec:exp-verifier}, consensus selection and the verifier select
from identical candidate pools. Differences in performance therefore come
only from the selection rule. Consensus selection performs better on all
geometry metrics without using a vision or reasoning model for selection. On
topology, we find no significant difference between topological consensus and
the verifier. On our data, the Euler characteristic may be too coarse to
provide a strong consensus signal.

Across all tested models and prompt variants, consensus selection improves over
random selection, but the gains vary. Consensus cannot help when all
candidates are identical. It can also fail when the same error appears in most
candidates because that error becomes part of the consensus. Moreover,
consensus favors candidates near the center of the pool and may miss a
high-quality outlier. This helps explain the gap to the oracle in
Figure~\ref{fig:scaling}. Future work could preserve or further
evaluate outliers alongside the consensus candidate.

Our criteria compare candidates at the level of the complete 3D model.
Geometric consensus normalizes and aligns the models before comparison.
Absolute dimensions therefore do not affect selection, even when they are
specified in the prompt. Small local features may also be hard to distinguish
with a global distance. To better capture such features, future work could
compare candidates at the feature or region level. Consensus selection could
also extend beyond geometry and topology. Other criteria could use dimensions,
manufacturability, or results from engineering simulations. Another
possibility is to create a new CAD model from features shared by several
candidates. This would combine candidates instead of selecting only one.
Another direction concerns how candidates are generated. We currently sample
all candidates in parallel. Whether consensus can also be formed across
sequential refinements of a single parametric CAD program remains open.

\section{Conclusion}
\label{sec:conclusion}

In this work, we introduced and comprehensively evaluated 3D CAD consensus selection for parametric CAD generation. The method
samples $N$ programs from an LLM, compiles them into 3D CAD models, and selects
the candidate that agrees most with the rest of the pool. Selection by
agreement is well established for text and code. In CAD, consensus can be
measured directly over properties of the compiled models. We use geometry and
topology, however, alternative CAD model properties, such as outcomes from engineering simulations, could be analogously integrated. On the exact candidate pools of a state-of-the-art
verifier-based CAD generation method, geometric consensus improves all
evaluated geometry metrics over the method's verifier. Regarding topology, we found no significant difference
between topological consensus and the verifier. Across all tested models and
prompt variant, geometric consensus also improves over random selection. Our
results show that the gains depend on the property used to define consensus.
The method is training-free and can be integrated into existing CAD generation
pipelines. CAD verifiers used for candidate selection should therefore be
compared against consensus selection on the same candidate pool.

\bibliographystyle{plainnat}
\bibliography{references}


\appendix

\newpage

\end{document}